\documentclass[prb,aps,twocolumn,showpacs]{revtex4}
\usepackage{epsfig}
\usepackage{amssymb}
\usepackage{amsmath}
\usepackage{amsfonts}

\begin{document}

\author{A.~Yu.~Kuntsevich$^{a}$, G.~M.~Minkov$^{b,c}$,
A.~A.~Sherstobitov$^{b,c}$, V.~M.~Pudalov$^a$}
\address{$^{a}$ P.\ N.\ Lebedev Physics Institute, 119991 Moscow, Russia \\
          $^{b}$ Institute of Metal Physics RAS, 620219 Ekaterinburg, Russia\\
          $^{c}$ Institute of Physics and Applied Mathematics, Ural State University, 620083 Ekaterinburg, Russia}

\title{Non-monotonic magnetoresistance of two-dimensional electron systems in the ballistic regime}

\begin{abstract}

We report experimental observations of a novel magnetoresistance
(MR) behavior of two-dimensional electron systems in perpendicular
magnetic field in the ballistic regime, for $k_BT\tau/\hbar>1$.
The MR grows with field and exhibits a maximum at fields
$B>1/\mu$, where $\mu$ is the electron mobility. As temperature
increases the magnitude of the maximum grows and its position
moves to higher fields. This effect is universal: it is observed
in various Si- and GaAs- based two-dimensional electron systems.
We compared our data with recent theory based on the Kohn anomaly
modification in magnetic field, and found qualitative similarities
and discrepancies.
\end{abstract}


\date{\today}

\maketitle

Two-dimensional (2D) degenerate electronic systems of high purity
($k_Fl\gg 1$) with isotropic energy spectrum are rather simple
objects, which seem to be  well understood. Within the classical
kinetic theory, \cite{landavshiz} the resistivity of such a system
should not depend on perpendicular magnetic field for
$\omega_c\tau<1$ (where $\omega_c=eB/m^*$ is the cyclotron
frequency, and $\tau$ -- the transport time). However, a
noticeable magnetoresistance (MR) is often seen in experiments
with  2D systems; such MR is usually attributed to quantum
corrections which are beyond the classical consideration. There
are two types of quantum corrections to conductivity: (i) weak
localization (WL), and (ii) electron-electron (e-e) interaction
correction (for a review, see Ref.~\cite{altshuler}). In the
diffusive regime ($k_B T \tau /\hbar \ll 1$,
$\tau/\tau_{\varphi}\ll 1$), both corrections give rise to
magnetoresistance with an amplitude proportional to
$\ln(T)$\cite{altshuler,gornyimirlin} whereas in the ballistic
regime ($k_BT\tau/\hbar>1$, $\tau/\tau_{\varphi}>1$) the
magnetoresistance should disappear\cite{gornyimirlin}. These
theoretical predictions for the MR have been verified in diffusive
and diffusive-to-ballistic crossover regimes
 in recent experiments
\cite{minkovballistictodiffusion,savchenko,kvonballistictodiffusion}
with 2D systems. A conventional belief (that the quantum
corrections to MR have to disappear at high temperatures) has made
the MR in purely ballistic regime out of the scope of experimental
interests. This theory prediction for the ballistic regime,
however, was not verified thoroughly. In order to shed light on
this issue, we measured MR in the ballistic regime with various
simple isotropic 2D electron systems. Contrary to the
 common belief we have found that the MR in perpendicular fields  does not vanish
at $k_BT\tau/\hbar
>1$; instead, it manifests a novel type of behavior:
{\em  the MR  depends non-monotonically on field and exhibits a
maximum, whose position scales with temperature for all samples}.

In this paper, we report observation and systematic studies of the
MR in the domain $k_Fl\gg1$, $k_BT\tau/\hbar>1$, where  the MR
should be missing. Experimentally, however, different Si-MOS
structures, GaAs/AlGaAs heterostructures and GaAs-based quantum
wells  were found to show a nonmonotonic MR. Our results provide
an evidence for a  universal origin of the effect. We compared our
data with a recent theory \cite{raikh} of e-e interaction
correction that employs modification of the Kohn anomaly by
magnetic field and did find some qualitative similarities.

 We used two Si-MOS samples (Si4, Si13
with peak mobilities 1-2 m$^2/$Vs) and  GaAs-AlGaAs
heterostructure 28, GaAs24 (mobility 24
m$^2/$Vs)\cite{pudalovgaas}, and gated quantum well structures
 AlGaAs-GaAs-AlGaAs (1520) and GaAs-InGaAs-GaAs
(3513) \cite{minkovballistictodiffusion}.  All samples were patterned as Hall bars.
Density of electrons in gated samples was
varied in situ. The relevant parameters of the samples, densities $n$~(in units of
$10^{12}$cm$^{-2}$), and mobilities
$\mu$~(m$^2/$Vs), are summarized in the following table:\\

\begin{tabular}{|c|c|c|c|c|c|}
  \hline
  Si- & $n$& $\mu$&GaAs-& $n$ & $\mu$ \\
  samples& $$&&samples&&\\
  \hline
  Si4 & 1.3 &  1.02 &  3513 & 1 & 2.2 \\
  Si4 & 1.7 &  1 & 28 & 0.35 & 24  \\
  Si4 & 2.35 & 0.96  &  24& 0.4&21\\
  Si4 & 3.4 &  0.93 & 1520 & 1.6 & 1.6  \\
  Si13 & 0.6 &  2.4 & 1520 & 1.4 & 1.5  \\
  Si13 & 0.7 & 2.3  & 1520 & 1  & 0.95  \\
  Si13 & 1 & 2.1  &1520 &0.8 &0.8 \\
  \hline
\end{tabular}

Samples were inserted into a cryostat with a superconducting
magnet; the field direction was always perpendicular to the 2D
sample plane. Temperature was varied in the range 1.3-60\,K. Both
components of the resistivity tensor were measured simultaneously
using the standard four-terminal technique with either SR-830
lock-in amplifier (samples Si4, Si13, 28, 24), or using
 rectangular current modulation (samples 1520,3513). Both, harmonic and
rectangular modulation was made at frequencies  12-33 Hz. Current
was chosen an order of 1\,$\mu$A,  to ensure the absence of
electron overheating.

\begin{figure}
\centerline{\psfig{figure=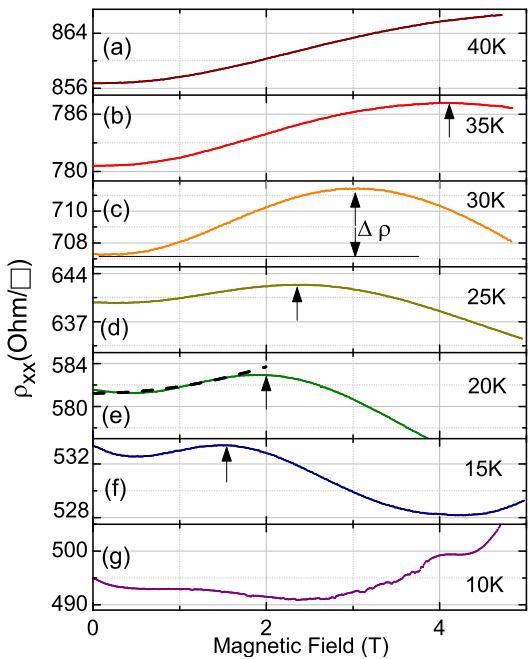,width=200pt}}
\begin{minipage}{3.2in}
\caption{(Color online)Magnetoresistance for sample Si4 at
different temperatures. Electron density
$n=1.72\cdot10^{12}$~cm$^{-2}$. Up-arrows mark positions of the
$\rho_{xx}$ maxima. $\Delta \rho$  designates the magnitude of the
MR. Dashed curve on the panel e shows fitting according to
Eq.~(\protect\ref{eq2}) with $\lambda^2=0.2$. $\hbar/k_B\tau
\approx 8$K.} \label{fig1}
\end{minipage}
\end{figure}

In order to exclude an admixture of the off-diagonal component of
the resistivity, we swept magnetic field from $-B$ to $B$, and
then symmetrized our data. Such a symmetrization is necessary for
reliable measurements of corrections to the resistivity whose
relative variations might be less than 1\%.

\begin{figure}
\centerline{\psfig{figure=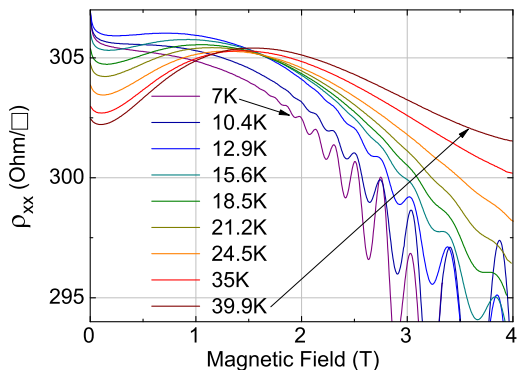,width=200pt}}
\begin{minipage}{3.2in}
\caption{(Color online)Magnetoresistance for sample 1520 at
different temperatures. Electron density
$n=1.4\cdot10^{12}$~cm$^{-2}$. Temperature values are indicated in
the figure. $\hbar/k_B\tau=13.5$K.} \label{fig2}
\end{minipage}
\end{figure}

Electron density values quoted in the paper were determined from
the slope of the Hall resistance versus $B$ as well as from the
period of Shubnikov - de Haas oscillations at  low temperatures.
Both results agreed with each other within 2$\%$. The highest
temperature in our experiments was chosen not to exceed 60\,K in
order the carrier density to remain constant and to avoid a
bypassing bulk conductivity.

Examples of our MR curves, obtained at different temperatures for
samples Si-4 and 1520 at fixed electron densities are shown in Figs.
\ref{fig1} and \ref{fig2}, respectively. As magnetic field is
increased from zero, at first, $\rho_{xx}$  sharply falls due to
weak localization suppression. Then $\rho_{xx}$  starts growing
and reaches a maximum
 at $B^{\rm max}$ field (indicated by the
arrows in Fig.~\ref{fig1}). After passing the maximum $\rho_{xx}$
decreases;  in higher fields, $|B|>1.5B^{\rm max}$, MR can become
either positive or negative depending on the sample, temperature,
electron density, etc. At the lowest temperatures, Shubnikov - de
Haas oscillations are seen in high fields, on top of the smooth
MR.

This unexpected nonmonotonic magnetoresistance is the main subject
of the current paper. We stress that this effect (i.e.,
nonmonotonic MR) is universal. The point is that in different
samples and at various electron densities it has similar features:
(i) MR is small (its typical magnitude  is less than 1\%), (ii)
the nonmonotonic MR is observed only for $T\geq1.3\hbar/k_B\tau$
\cite{usednotations}, (iii) the MR maximum grows in magnitude and
moves to higher magnetic fields as temperature increases (the
position of the maximum exceeds $\omega_c\tau>1$  and is roughly
proportional to $T$).

Comparing the data from Figs.~(\ref{fig1}) and Fig.~(\ref{fig2})
for Si-MOSFET and GaAs QW-samples with similar mobilities and
densities, we see that the MR takes a maximum at similar
temperatures and magnetic fields, and at similar $\omega_c\tau$
values. This result indicates that the MR has an orbital rather
than spin origin because the Zeeman energies $g^*\mu_B$ differ by
a factor of 5 for these two different material systems. Also, this
effect has nothing to do with WL and e-e-interaction diffusive
corrections \cite{altshuler} because it survives at such high
temperatures as $k_BT\tau/\hbar \approx 20$ for samples 28 and 24
at $T=20$K.

Searching for possible semiclassical mechanisms, we have to note
that most of the theoretical models for the case of short-range
scatterers \cite{short-range} predict a {\em negative, monotonic
and temperature independent} magnetoresistance, due to the memory
effects \cite{dmitriev}. A positive, though $T$-independent,
magnetoresistance was predicted in Ref.~\cite{mirlin}, due to
non-markovian scattering. The latter type of MR  was
experimentally observed  in very clean samples and for classically
large magnetic fields \cite{quasiclassicexperiment},
$\omega_c\tau\gg 1$. Therefore, we
 conclude that the aforecited semiclassical mechanisms can't explain the nonmonotonic MR
observed in our experiments.

Recently, Sedrakyan and Raikh~\cite{raikh}  suggested a new MR
mechanism, which causes a maximum of resistivity in not-too-strong
magnetic fields $\omega_c\tau\sim 1$ in the ballistic regime
($k_BT\tau/\hbar>1$). This new mechanism seems to give the best
starting point for comparison with our measurements. The MR in
Ref.~\cite{raikh} originates from the e-e interaction correction
to conductivity. According to Ref.~\cite{ZNA}, e-e interaction
 corrections to conductivity  arise from scattering of electrons on Fridel's
oscillations of electron density around impurities. Fridel's
oscillations are a manifestation of the Kohn $2k_F$ anomaly in
screening. Magnetic field applied perpendicular to the 2D
plane modifies the electron spectrum and the Kohn anomaly; hence,
the field affects screening and electron scattering. In Ref.~
\cite{raikh} this point was taken into account and shown to give
rise to the second-order correction in the ballistic regime (see
Eq.~(5) from Ref. \cite{raikh}):
\begin{equation}
\frac{\delta\sigma_{xx}}{\sigma_{xx}}=4\lambda^2\left(\frac{\pi
k_BT}{E_F}\right)^{3/2}F_2\left(\frac{\omega_c
E_F^{1/2}}{2\pi^{3/2}(k_B T)^{3/2}} \right) \label{eq1},
\end{equation}
where $\lambda=1+3F_0^{\sigma}/(1+F_0^{\sigma})$ is the
interaction parameter \cite{commentonlambda}.

Several predictions can be made based on this equation: (1) the
correction to resistivity in small fields is always positive, (2)
 $(\delta\sigma_{xx}/\sigma_{xx})\cdot
\left(E_F/T\right)^{3/2}$ is a universal function of
$\omega_cE_F^{1/2}/T^{3/2}$ for a given interaction strength
$\lambda$, and (3) MR has a maximum at
$\omega_c\tau\approx1/\sqrt{3}$.

\begin{figure}[h]
\centerline{\psfig{figure=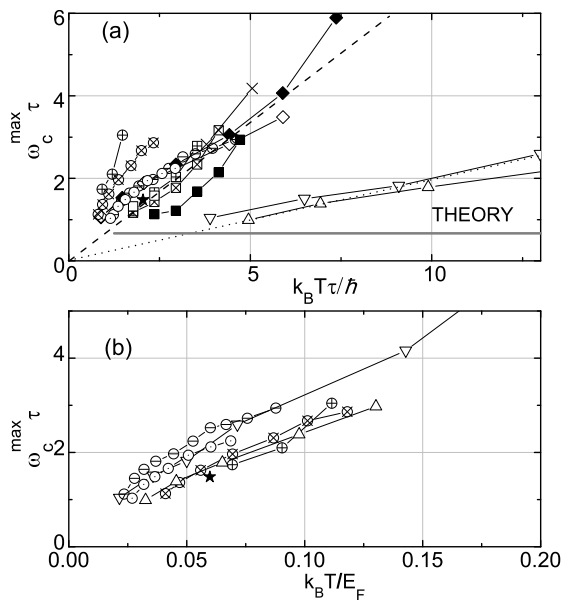,width=210pt}}
\begin{minipage}{3.2in}
\caption{(a) $\omega_c^{\rm max}\tau$ value versus dimensionless
temperature $k_BT\tau/\hbar$ for all samples. Electron densities
(in units of $10^{12}$ cm$^{-2}$) are $\diamondsuit $ - $n=0.6$
(Si13);$\blacklozenge$ - $n=0.7$ (Si13);$\times$ - $n=1$ (Si13);
$\blacksquare$ - $n=1.3$ (Si4); $\boxtimes$ - $n=1.7$ (Si4);
$\boxplus$ - $n=2.35$ (Si4);$\Box$ - $n=3.4$ (Si4); $\odot$ -
$n=1.4$ (1520); $\otimes$ - $n=1$ (1520); $\ominus$ - $n=1.6$
(1520); $\oplus$ - $n=0.8$ (1520);$\star$ - $n=1$ (1520);
$\triangledown$ - $n=0.35$ (28); $\triangle$ - $n=0.4$ (24).
Dashed line corresponds to $\hbar\omega_c^{\rm max}=0.7k_BT$.
Dotted line is $\hbar\omega_c^{\rm max}=0.2k_BT$. Horizontal thick
line is the theoretical prediction (see in the text). (b) The same
data for GaAs-based samples solely scale in coordinates
$\omega_c^{\rm max}\tau$ versus $k_BT/E_F$. } \label{fig3}
\end{minipage}
\end{figure}

By comparing these theoretical predictions with our data, we find
that prediction (1) is always fulfilled after subtraction of the
weak localization. As for prediction (2), the $\rho_{xx}(B)$-data
for different temperatures and over the whole range of magnetic
fields do not scale as the theory predicts. Furthermore, the
position of the MR maximum in our data is temperature dependent
and corresponds to $\omega_c\tau\approx 1-3$, contrary to
prediction (3). Moreover, the magnitude of the MR falls as
temperature raises in the theory, whereas {\em in our experiment
it grows with temperature}. Evidently, there is no complete
agreement between the theory \cite{raikh} and our experiment. We
note finally, that the magnitude of the MR maximum $\Delta \rho$
seems do not scale  with any dimensionless combination of $k_BT$,
$\hbar/\tau$, $\hbar\omega_c^{\rm max}$ and $E_F$. This is also in
contrast with the theory, where $\Delta \rho$ should be $\propto
\hbar^2/[\tau^2E_F^{0.5}(k_BT)^{1.5}]$.

 According to Eq.\ref{eq1}, the magnitude of the effect is proportional
to the interaction constant $\lambda^2$. Therefore, one could
estimate $\lambda^2$ value from the experimental data. In the
theory, the maximum of MR inevitably results from
$(1-\omega_c^2\tau^2)$ prefactor in resistivity tensor  and should
occur at $\omega_c\tau\approx 1/\sqrt{3}$. On the other hand, in
the experiment the $\rho_{xx}$ maximum is  always observed at
$\omega_c\tau>1$, which indicates that this prefactor is weaker
than in the theory.
Therefore, for the order-of-magnitude comparison, 
we rewrite Eq.~(\ref{eq1}) for resistivity and omit the
$[1-(\omega_c\tau)^2]$ prefactor:
\begin{equation}
\frac{\delta\rho_{xx}}{\rho_{xx}}=-4\lambda^2\left(\frac{\pi
k_BT}{E_F}\right)^{3/2}F_2\left(\frac{\omega_c
E_F^{1/2}}{2\pi^{3/2}(k_B T)^{3/2}} \right). \label{eq2}
\end{equation}
Example of the   corresponding fitting with a single variable
parameter $\lambda^2$ is shown in Fig.~\ref{fig1}\,e. The fit was
performed in the limited range of magnetic fields
$0.15\omega_c^{\rm max}<\omega_c<0.65\omega_c^{\rm max}$, i.e. in
the range of the applicability of Eq.~(\ref{eq2}) which ignores
weak localization  and  the  MR maximum. The $\lambda^2$  values
obtained from the fit  appeared to be temperature dependent i.e.
grew monotonically from 0.1-0.4 to 1-3 as temperature was
increased from $1.3\hbar/(k_B\tau)$ to maximal temperature. Surely
this temperature dependence causes the lack of the  scaling
predicted by Eq.~(\ref{eq1}). Moreover, $\lambda^2$-values
obtained from the fitting  don't show a systematic dependence on
carrier density and on material system.

On the other hand, the  $\lambda^2$ value in our range of
densities may be evaluated from the earlier measurements of
$F_0^\sigma(n)$ parameter. The calculated $\lambda$ values are
$T$-independent and lie in the interval from 0.2 to 0.5 for
GaAs-based structures \cite{minkovballistictodiffusion} and from
1.5 to 5 for Si-based structures\cite{lambdaestimates}. We
conclude therefore that the observed MR  disagrees qualitatively
with the theory, though the theory predicts the MR of the right
order of magnitude.

In Fig.~\ref{fig3}\,a, the position of the MR maximum of
temperature. The $\omega_c^{\rm max}\tau$ value systematically
exceeds the theoretical expectation $1/\sqrt{3}$ (horizontal thick
line in Fig. \ref{fig3}\,a) and approximately equals
$0.7k_BT\tau/\hbar$ for most of the data (dashed curve in
Fig.~\ref{fig3}a). For samples with the highest mobility (24,28),
the slope $\omega_c^{\rm max}\tau/(k_BT\tau/\hbar) \approx 0.2$
whereas for GaAs-based sample with the lowest mobility the slope
exceeds 0.7. In order to take this fact into account we have
applied another scaling, in coordinates versus $k_BT/E_F$ (see
Fig.~\ref{fig3}b). It is remarkable, that for GaAs-based 2D
systems with mobilities and conductivities ranging by more than an
order of magnitude, the $\omega_c^{\rm max}\tau$  data indeed
scale reasonably, the result that might suggest a clue for
understanding the effect.

The data for Si-based structures are not shown in Fig.\ref{fig3}b
because they fall out of the $T/E_F$ scaling. In order to
understand the origin of the difference in scaling for Si- and
GaAs- samples, we note that for GaAs-based samples in high fields,
$B>B^{\rm max}$, the MR is always negative while for Si-based
samples it can be either negative or positive, depending on
particular sample and electron density. It means that some other
mechanisms affect MR in Si-MOSFETs in strong perpendicular fields
$B>B^{\rm max}$ and shift the MR maximum. It is also worthy of
noting that in Si the discussed  weak MR  is observed at such high
temperatures where metallic temperature dependence of the
resistivity is strong and nonlinear with respect to $T$, and
hence, the first order interaction corrections \cite{ZNA} are
inapplicable.

We note also, that due to clear reasons
the nonmonotonic MR in the ballistic regime  $T\geq1.3\hbar/k_B\tau$ was not oserved
 in the following cases:
(i) Si-MOSFETs in the domain of strong interactions ($n<6\cdot 10^{11}$cm$^{-2}$)  where the giant negative MR developes and dominates over other weak effects \protect\cite{voiskovskii},
(ii) Si-MOSFETs for such high temperatures where  Fermi-gas is non-degenerate ($T/E_F\geq 0.5$),
(iii) GaAs based samples at such high temperatures that the carrier density becomes $B$- and $T$-dependent.

{\bf Conclusions.} In this paper we report experimental
observation of the novel non-monotonic behavior of the magnetoresistance
 for 2D electron systems in perpendicular field. This MR is  intrinsic to various
2D systems (Si-MOSFETs,  GaAs and InGaAs quantum wells, and
GaAs/AlGaAs-heterostructures) and occurs  in the ballistic regime
of high temperatures  $T\tau > 1$. The MR is positive in low
fields and  reaches a maximum at $\omega_c\tau=1-3$; the position
of the maximum $\omega_c^{\rm max}$ scales linearly with
temperature for all samples. We compare our data with recently
suggested MR mechanism \cite{raikh} and find some similarities:
(i) the MR is always positive  in low field, (ii) the MR exhibits
a maximum in higher field and (iii) the MR magnitude is of the
same  order of magnitude as predicted. However, other features of
our experimental data are in discrepancy with the theory
Ref.~\cite{raikh}: (i) the MR maximum is achieved in fields which
are noticeably higher than predicted, (ii) the position of the MR
maximum linearly depends on temperature rather than remains
constant, (iii) the magnitude of the effect increases with
temperature rather than decreases, as predicted.

Some clue to understanding the effect may be provided by  scaling
of the MR maximum position $\omega_c^{\rm max}\tau$ versus
$T/T_F$,  which is empirically observed for various GaAs-samples
in wide ranges of temperature, density and mobility. The
 observation of the nonmonotonic MR shows that the magnetotransport theory is still incomplete, at
least for the ballistic regime, and requires further
consideration.

{\bf Acknowledgements}. We thank M.~E.~Raikh, T.~A.~Sedrakyan, and
I.~S.~Burmistrov for discussions and I.~E.~Bulyzhenkov for valuable comments.
The work was supported by RFBR, Programs of the
RAS, Russian Ministry for Education and Science, and the Program
``Leading Scientific Schools''.


\begin{references}
\bibitem{usednotations}Throughout this paper the following notations are used:
$E_F=p_F^2/2m^*, \omega_c=eB/m^*, \tau=\sigma_Dm^*/ne^2$, where
$m^*$ is a band mass of the electron ($0.067m_e$ for GaAs and
$0.21m_e$ for Si) $\sigma_D$ is the Drude conductivity. The latter is found by
extrapolating the conductivity in ballistic regime to zero
temperature.
\bibitem{short-range} For short range scatterers case, $\tau$ should be of the same order as the all-angle scattering time $\tau_q$. From temperature  dependence  of the
Shubnikov-de Haas oscillations amplitude we found
that for our samples $\tau_q$ approximately equals $\tau$, indicating short-range
scattering.
\end{references}
\end{document}